\def \seclogo{}

\documentclass[article]{ieeetj}

\usepackage{amsmath,graphicx,hyperref,xcolor,amsfonts,multicol,placeins}

\hypersetup{
    colorlinks=true,
    citecolor=blue,
    linkcolor=blue,
    filecolor=magenta,      
    urlcolor=blue,
}
\usepackage{booktabs}
\graphicspath{{figures/}} 

\begin{document}

\title{ICASSP 2023 Acoustic Echo Cancellation Challenge}
\author{Ross Cutler, Ando Saabas, Tanel P\"arnamaa, Marju Purin, Evgenii Indenbom, Nicolae-C\u{a}t\u{a}lin Ristea, Jegor Gužvin, Hannes Gamper, Sebastian Braun, Robert Aichner}
\affil{Microsoft Corporation, Redmond, USA}
\corresp{Corresponding author: Ross Cutler (email:ross.cutler@microsoft.com).}


\begin{abstract}
The ICASSP 2023 Acoustic Echo Cancellation Challenge is intended to stimulate research in acoustic echo cancellation (AEC), which is an important area of speech enhancement and is still a top issue in audio communication. This is the fourth AEC challenge and it is enhanced by adding a second track for personalized acoustic echo cancellation, reducing the algorithmic + buffering latency to 20ms, as well as including a full-band version of AECMOS \cite{purin_aecmos_2022}. We open source two large datasets to train AEC models under both single talk and double talk scenarios. These datasets consist of recordings from more than 10,000 real audio devices and human speakers in real environments, as well as a synthetic dataset. We open source an online subjective test framework and provide an objective metric for researchers to quickly test their results. The winners of this challenge were selected based on the average mean opinion score (MOS) achieved across all scenarios and the word accuracy (WAcc) rate.
\end{abstract}

\maketitle

\section{Introduction}
\label{sec:intro}
With the growing popularity and need for working remotely, the use of teleconferencing systems such as Microsoft Teams, Skype, WebEx, Zoom, etc., has increased significantly. It is imperative to have good quality calls to make the user's experience pleasant and productive. The degradation of call quality due to acoustic echoes is one of the major sources of poor speech quality ratings in voice and video calls. While digital signal processing (DSP) based AEC models have been used to remove these echoes during calls, their performance can degrade when model assumptions are violated, e.g., fast time-varying acoustic conditions, unknown signal processing blocks or non-linearities in the processing chain, or failure of other models (e.g., background noise estimates). This problem becomes more challenging during full-duplex modes of communication where echoes from double talk scenarios are difficult to suppress without significant distortion or attenuation \cite{noauthor_ieee_2010}. 

With the advent of deep learning techniques, many supervised learning algorithms for AEC have shown better performance compared to their classical counterparts, e.g., \cite{fazel_cad-aec_2020, halimeh_efficient_2020, ma_acoustic_2020}. Some studies have also shown good performance using a combination of classical and deep learning methods such as using adaptive filters and recurrent neural networks (RNNs) \cite{ma_acoustic_2020, zhang_deep_2019} but only on synthetic datasets. While these approaches are promising, they lack evidence of their performance on real-world datasets with speech recorded in diverse noise and reverberant environments. This makes it difficult for researchers in the industry to choose a good model that can perform well on a representative real-world dataset.

Most AEC publications use objective measures such as echo return loss enhancement (ERLE) \cite{enzner_acoustic_2014} and perceptual evaluation of speech quality (PESQ) \cite{noauthor_itu-t_2000}. ERLE in dB is defined as: 

\begin{equation}
\text{ERLE} = 10\log_{10} \frac{\mathbb{E}[y^2(n)]}{\mathbb{E}[e^2(n)]} 
\end{equation}

\noindent where $y(n)$ is the microphone signal, and $e(n)$ is the residual echo after cancellation. ERLE is only appropriate when measured in a quiet room with no background noise and only for single talk scenarios (not double talk), where we can use the processed microphone signal as an estimate for $e(n)$. PESQ has also been shown to not have a high correlation to subjective speech quality in the presence of background noise \cite{avila_non-intrusive_2019}. Using the datasets provided in this challenge we show that ERLE and PESQ have a low correlation to subjective tests (Table \ref{tab:correlation}). In order to use a dataset with recordings in real environments, we can not use ERLE and PESQ. A more reliable and robust evaluation framework is needed that everyone in the research community can use, which we provide as part of the challenge.

\begin{table}
    \centering
    \caption{Pearson Correlation Coefficient (PCC) and Spearman's Rank Correlation Coefficient (SRCC) between objective and subjective P.808 results on single talk echo scenarios (see Section \ref{sec:framework}).}
    \label{tab:correlation}
    \begin{tabular}{ccc}
    \toprule
    {} & PCC & SRCC \\
    \midrule
    ERLE & 0.31 & 0.23 \\
    PESQ & 0.67 & 0.57 \\
    \bottomrule
    \end{tabular}
\end{table}

This AEC challenge is designed to stimulate research in the AEC domain by open-sourcing a large training dataset, test set, and subjective evaluation framework. We provide two new open-source datasets for training AEC models. The first is a real dataset captured using a large-scale crowdsourcing effort. This dataset consists of real recordings that have been collected from over 10,000 diverse audio devices and environments. The second dataset is synthesized from speech recordings, room impulse responses, and background noise derived from \cite{reddy_interspeech_2020}. An initial test set was released for the researchers to use during development and a blind test set near the end, which has been used to decide the final competition winners. We believe these datasets are large enough to facilitate deep learning and representative enough for practical usage in shipping telecommunication products (e.g., see \cite{indenbom_deepvqe_2023}).

This is the fourth AEC challenge we have conducted. The first challenge was held at ICASSP 2021 \cite{sridhar_icassp_2021}, the second at INTERSPEECH 2021 \cite{cutler_interspeech_2021}, and the third at ICASSP 2022 \cite{cutler_icassp_2022}. These challenges had 49 participants with entries ranging from pure deep models, hybrid linear AEC + deep echo suppression, and DSP methods. While the submitted AECs have consistently been getting better, there is still significant room for improvement as shown in Table \ref{tab:improvement}. The two largest areas for improvement are (1) Single Talk Near End quality, which is affected by background noise, reverberation, and capture device distortions, and (2) Double Talk Other Degradations, which includes missing audio, distortions, and cut-outs.  In addition, the overall challenge metric, $M$ was 0.883 out of 1.0 in the ICASSP 2022 challenge, which also shows significant room for improvement. 

\begin{table}[]
    \centering
    \caption{Amount of improvement remaining based on the ICASSP 2022 AEC Challenge \cite{cutler_icassp_2022}.}
    \begin{tabular}{cc}
    \toprule
        Area & Headroom \\
        \hline 
        Single talk near end MOS & 0.60 \\
        Single talk far end MOS & 0.19 \\
        Double talk echo & 0.25 \\
        Double talk other & 0.57 \\
        WAcc & 0.19 \\
    \bottomrule
    \end{tabular}
    \label{tab:improvement}
    
\end{table}

To improve the challenge and further stimulate research in this area we have made the following changes:

\begin{itemize}
    \item We included a second track for personalized AEC. Based on the excellent results for personalized noise suppression in the ICASSP 2022 Deep Noise Suppression Challenge \cite{dubey_icassp_2022}, we expected significant improvements for the double talk scenario.
    \item We reduced the algorithmic latency + buffering latency from 40ms to 20ms which is necessary for use in real-time collaboration systems. This will make getting the same speech quality done in previous challenges more difficult. 
    \item We provided a full-band version of AECMOS so it can be better used for full-band training and testing. AECMOS is freely available at \url{https://github.com/microsoft/AEC-Challenge}.
\end{itemize}

An overview of the four AEC challenges is given in Table \ref{tab:challenges}.

\begin{table*}
    \centering
    \caption{Summary of AEC challenges. BAK and SIG are measurements of the background noise quality and speech signal quality.}
    \begin{tabular}{lllll}
    \toprule
        Challenge & Tracks & Datasets & Algorithmic +  & Notes \\
        & & & Buffering Latency & \\
        \hline 
        ICASSP 2021	& Real-time & 2,500 real environments & 40ms & Crowdsourced P.831 \\
        & & Synthetic & & \\
        INTERSPEECH 2021 & Real-time & 5,000 real environments & 40ms & Made test set more comprehensive \\
        & Non-real-time & Synthetic & & Increased subjective test framework accuracy \\
        & & & & Added AECMOS service \\
        ICASSP 2022 & Real-time	& 7,500 real environments & 40ms & Added mobile scenarios\\
        & & Synthetic & & Added WAcc \\
        & & & & Made datasets, test sets full band \\
        ICASSP 2023	& Real-time & 10,000 real environments & 20ms & Added fullband AECMOS \\
        & Personalized & Synthetic & & Split near end quality into BAK and SIG \\
    \bottomrule
    \end{tabular}
    \label{tab:challenges}
\end{table*}

Related work is reviewed in Section \ref{sec:related_work}. The challenge description is given in Section \ref{sec:challenge}. The training dataset is described in Section \ref{sec:data}, and the test set in Section \ref{sec:data_test}. We describe a baseline deep neural network-based AEC method in Section \ref{sec:model}. The online subjective evaluation framework is discussed in Section \ref{sec:framework}, and the objective function in Section \ref{sec:metric}. The challenge metric is given in Section \ref{sec:challenge_metric} and the challenge rules are described in \url{https://aka.ms/aec-challenge}. The results and analysis are given in Section \ref{sec:results}, and conclusions are discussed in Section \ref{sec:conclusions}.

\section{Related work}
\label{sec:related_work}
There are many standards for measuring AEC performance. For objective metrics, IEEE 1329 \cite{noauthor_ieee_2010} defines metrics like terminal coupling loss for single talk (TCLwst) and double talk (TCLwdt), which are measured in anechoic chambers. TIA 920 \cite{noauthor_tia-920_2002} uses many of these metrics but defines required criteria. ITU-T Rec. G.122 \cite{noauthor_itu-t_2012-3} defines AEC stability metrics, and ITU-T Rec. G.131 \cite{noauthor_itu-t_2003-2} provides a useful relationship of acceptable Talker Echo Loudness Rating and one-way delay time. ITU-T Rec. G.168 \cite{noauthor_itu-t_2012-1} provides a comprehensive set of AEC metrics and criteria. However, it is not clear how to combine these dozens of metrics into a single metric, or how well these metrics correlate to subjective quality. 

Subjective speech quality assessment is the gold standard for evaluating speech enhancement processing and telecommunication systems, and the ITU-T has developed several recommendations for subjective speech quality assessment. ITU-T P.800 \cite{noauthor_itu-t_1996} describes lab-based methods for the subjective determination of speech quality. In P.800, users are asked to rate the quality of speech clips on a Likert scale from 1: Poor to 5: Excellent. Many ratings are taken for each clip, and the average score for each clip is the MOS. ITU-T P.808 \cite{noauthor_itu-t_2018} describes a crowdsourcing approach for conducting subjective evaluations of speech quality. It provides guidance on test material, experimental design, and a procedure for conducting listening tests in the crowd. The methods are complementary to laboratory-based evaluations described in P.800. An open-source implementation of P.808 is described in \cite{naderi_open_2020}. ITU-T P.835 \cite{noauthor_itu-t_2003} provides a subjective evaluation framework that gives standalone quality scores of speech (SIG) and background noise (BAK) in addition to the overall quality (OVRL). An open-source implementation of P.835 is described in \cite{naderi_subjective_2021}. More recent multidimensional speech quality assessment standards are ITU-T P.863.2 \cite{noauthor_itu-t_2022} and P.804 \cite{noauthor_itu-t_2017-1} (listening phase), which measure noisiness, coloration, discontinuity, and loudness. An open-source implementation of P.804 using crowdsourcing is described in \cite{naderi_multi-dimensional_2023}. 

ITU-T Rec. P.831 \cite{noauthor_itu-t_1998-1} provides guidelines on how to conduct subjective tests for network echo cancellers in the laboratory. ITU-T Rec. P.832 \cite{noauthor_itu-t_2000} focuses on the hands-free terminals and covers a broader range of degradations. Cutler et al.~\cite{cutler_crowdsourcing_2021} provide an open-source crowdsourcing tool extending P.831 and P.832 and include validation studies that show it is accurate compared to expert listeners and repeatable across multiple days and different raters.  Purin et al.~\cite{purin_aecmos_2022} created an objective metric, AECMOS, based on this tool's results on hundreds of different AEC models. AECMOS has a high correlation to subjective opinion. 

While there have been hundreds of papers published on deep echo cancellation since the first AEC challenge, we feel the winners of each challenge are of special note since they have been tested and evaluated using realistic and challenging test sets and subjective evaluations. Table \ref{tab:challenge_winners} provides the top three papers for each previous AEC challenge. Note that because the performance rankings and paper acceptances were decoupled in ICASSP 2021 and INTERSPEECH 2021, the challenge placement and performance rankings are not identical, and for INTERSPEECH 2021 not well correlated. For ICASSP 2022 and 2023, the top five papers based on the challenge performance were submitted for review, fixing the disparity between paper acceptance and model performance. 

\begin{table}[tb]
\centering
\caption{AEC Challenge top 3 performers.}
\label{tab:challenge_winners}
\begin{tabular}{l c c  c c  c c} 
\toprule
Challenge & 1st & Rank & 2nd & Rank & 3rd & Rank \\
\midrule
ICASSP 2021 & \cite{valin_low-complexity_2021} & 1 & \cite{wang_weighted_2021} & 2 & \cite{westhausen_acoustic_2020} & 5 \\ 
INTERSPEECH 2021 & \cite{peng_acoustic_2021} & 6 & \cite{zhang_f-t-lstm_2021} & 8 & \cite{pfeifenberger_acoustic_2021} & 10 \\
ICASSP 2022 & \cite{zhang_multi-scale_2022} & 1 & \cite{zhao_deep_2022} & 2 & \cite{zhang_multi-task_2022} & 3 \\
\bottomrule
\end{tabular}
\end{table}

\section{Challenge description}
\label{sec:challenge}
\subsection{Tracks}
This challenge included two tracks:

\begin{itemize}
    \item Non-personalized AEC. This is similar to the ICASSP 2022 AEC Challenge.
    \item Personalized AEC. This adds speaker enrollment for the near end speaker. A speaker enrollment is a 15-25 second recording of the near end speaker that can be used for adopting the AEC for personalized echo cancellation. For training and model evaluation, the datasets in \url{https://github.com/microsoft/AEC-Challenge}  can be used, which include both echo and near end only clips from users. For the blind test set, the enrollment clips will be provided.
\end{itemize}

\subsection{Latency and runtime requirements}
Algorithmic latency is defined by the offset introduced by the whole processing chain including short time Fourier transform (STFT), inverse STFT, overlap-add, additional lookahead frames, etc., compared to just passing the signal through without modification. It does not include buffering latency. Some examples are:

\begin{itemize}
    \item A STFT-based processing with window length = 20 ms and hop length = 10 ms introduces an algorithmic delay of window length – hop length = 10 ms.
    \item A STFT-based processing with window length = 32 ms and hop length = 8 ms introduces an algorithmic delay of window length – hop length = 24 ms.
    \item An overlap-save-based processing algorithm introduces no additional algorithmic latency.
    \item A time-domain convolution with a filter kernel size = 16 samples introduces an algorithmic latency of kernel size – 1 = 15 samples. Using one-sided padding, the operation can be made fully “causal”, i.e., a left-sided padding with kernel size-1 samples would result in no algorithmic latency.
    \item A STFT-based processing with window\_length = 20 ms and hop\_length = 10 ms using 2 future frames information introduces an algorithmic latency of (window\_length – hop\_length) + 2*hop\_length = 30 ms.
\end{itemize}

Buffering latency is defined as the latency introduced by block-wise processing, often referred to as hop length, frame-shift, or temporal stride. Some examples are:

\begin{itemize}
    \item A STFT-based processing has a buffering latency corresponding to the hop size.
    \item A overlap-save processing has a buffering latency corresponding to the frame size.
    \item A time-domain convolution with stride 1 introduces a buffering latency of 1 sample.
\end{itemize}

Real-time factor (RTF) is defined as the fraction of time it takes to execute one processing step. For a STFT-based algorithm, one processing step is the hop size. For a time-domain convolution, one processing step is 1 sample. RTF = compute time / time step.

All models submitted to this challenge must meet all of the below requirements:

\begin{enumerate}
    \item To be able to execute an algorithm in real-time, and to accommodate for variance in compute time which occurs in practice, we require RTF $\leq$ 0.5 in the challenge on an Intel Core i5 Quadcore clocked at 2.4 GHz using a single thread.
    \item Algorithmic latency + buffering latency $\leq$ 20ms.
    \item No future information can be used during model inference.
\end{enumerate}

\section{Training datasets}
\label{sec:data}
The challenge includes two open-source datasets, one real and one synthetic. The datasets are available at \url{https://github.com/microsoft/AEC-Challenge}.

\subsection{Real dataset}
\label{ssec:real_data}
 The first dataset was captured using a large-scale crowdsourcing effort. This dataset consists of more than 50,000 recordings from over 10,000 different real environments, audio devices, and human speakers in the following scenarios:

\begin{enumerate}
    \item Far end single talk, no echo path change
    \item Far end single talk, echo path change
    \item Near end single talk, no echo path change
    \item Double talk, no echo path change
    \item Double talk, echo path change
    \item Sweep signal for RT60 estimation
\end{enumerate}

\begin{figure}[t]
    \centering
    \includegraphics[width=160pt]{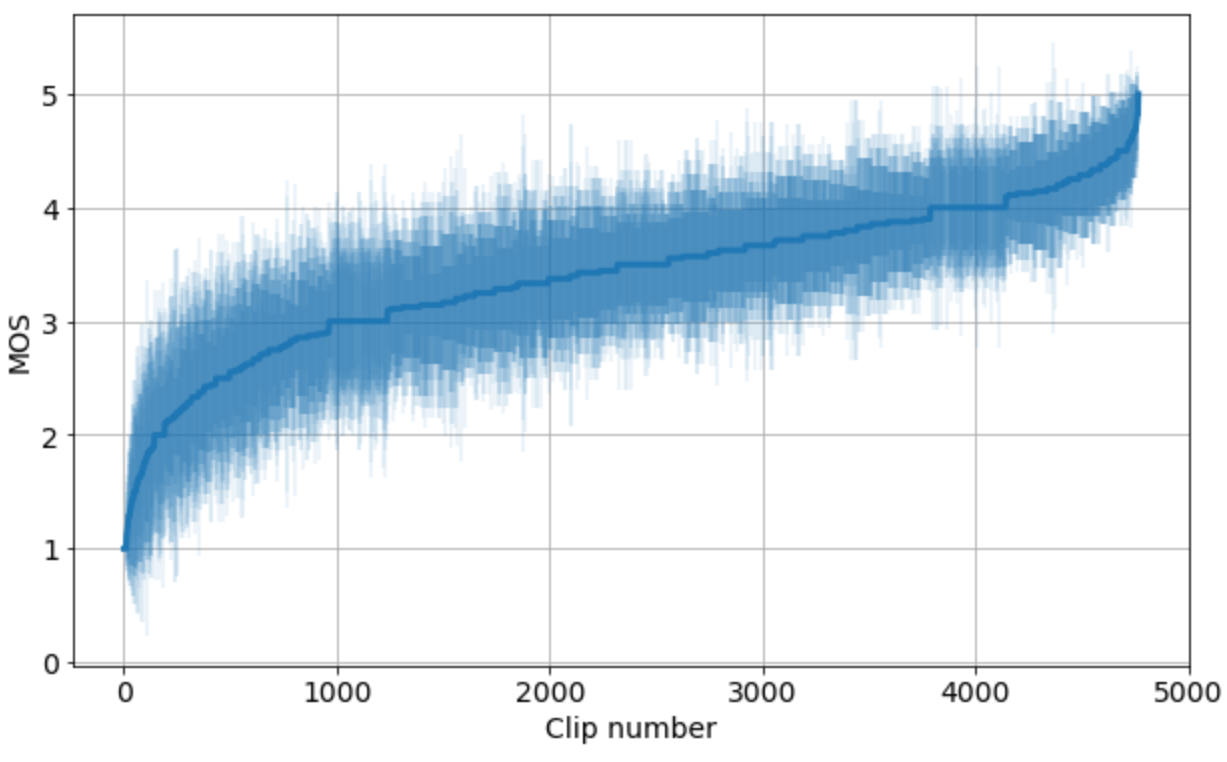}
    \caption{Sorted near end single talk clip quality (P.808) with 95\% confidence intervals.}
    \label{fig:nearend}
\end{figure}

\begin{figure}
    \centering
    \includegraphics{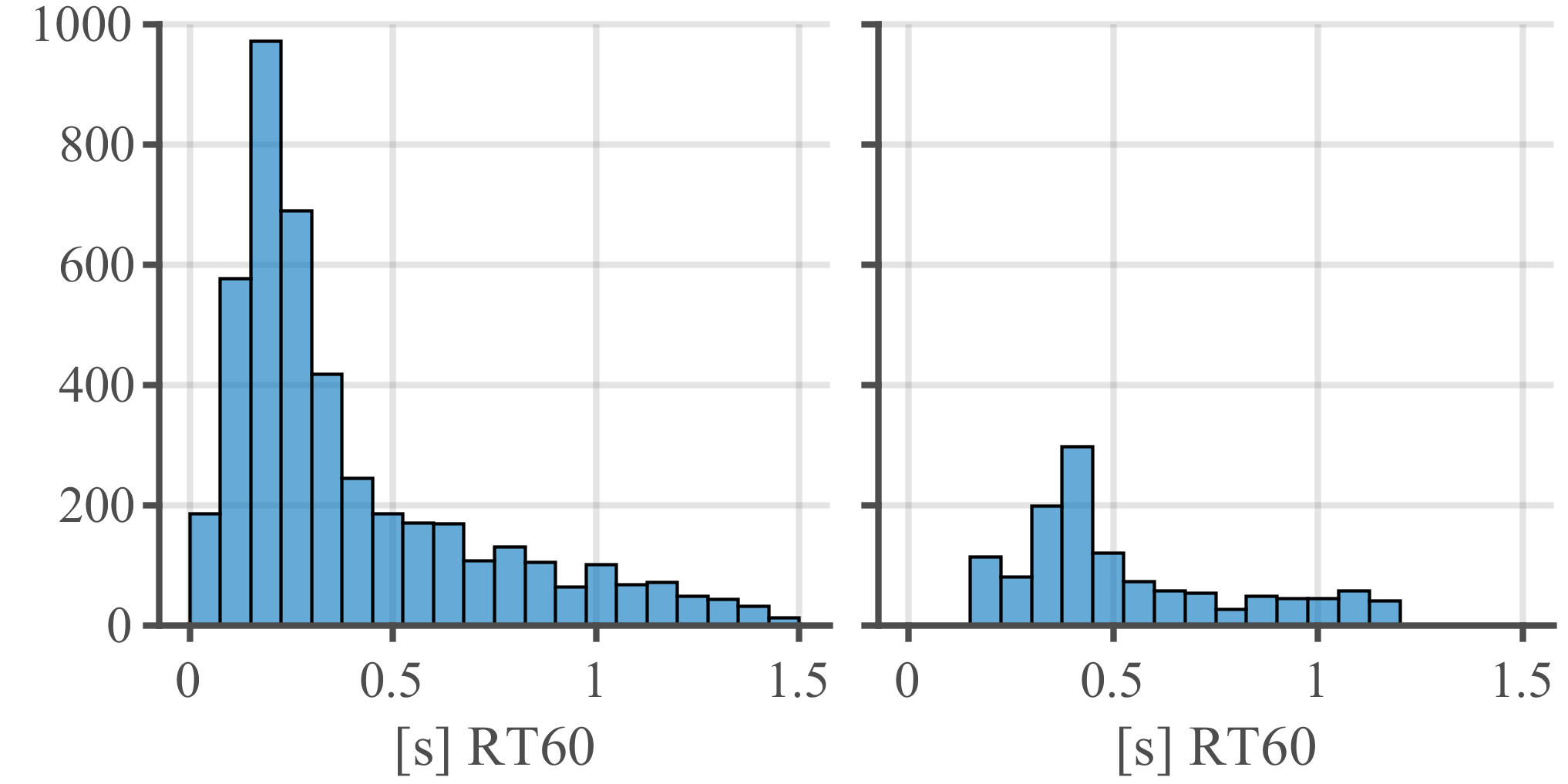} 
    \caption{Distribution of reverberation time (RT60) for Desktop (left) and Mobile (right).}
    \label{fig:rt60}
\end{figure}

RT60 is the time for an initial signal's sound pressure level to attenuate 60 dB from its original level. For the far end single talk case, there is only the loudspeaker signal (far end) played back to the users and users remain silent (no near end speech). For the near end single talk case, there is no far end signal and users are prompted to speak, capturing the near end signal. For double talk, both the far end and near end signals are active, where a loudspeaker signal is played and users talk at the same time. Echo path changes were incorporated by instructing the users to move their device around or bring themselves to move around the device.  The RT60 distribution for 4387 desktop environments in the real dataset for which impulse response measurements were available is estimated using a method by Karjalainen et al.~\cite{karjalainen_estimation_2001} and shown in Figure \ref{fig:rt60}. For 1251 mobile environments the RT60 distribution shown was estimated blindly from speech recordings~\cite{gamper_blind_2018}. The RT60 estimates can be used to sample the dataset for training. The near end single talk speech quality is given in Figure \ref{fig:nearend}.

\begin{figure}
    \centering
    \includegraphics[width=200pt]{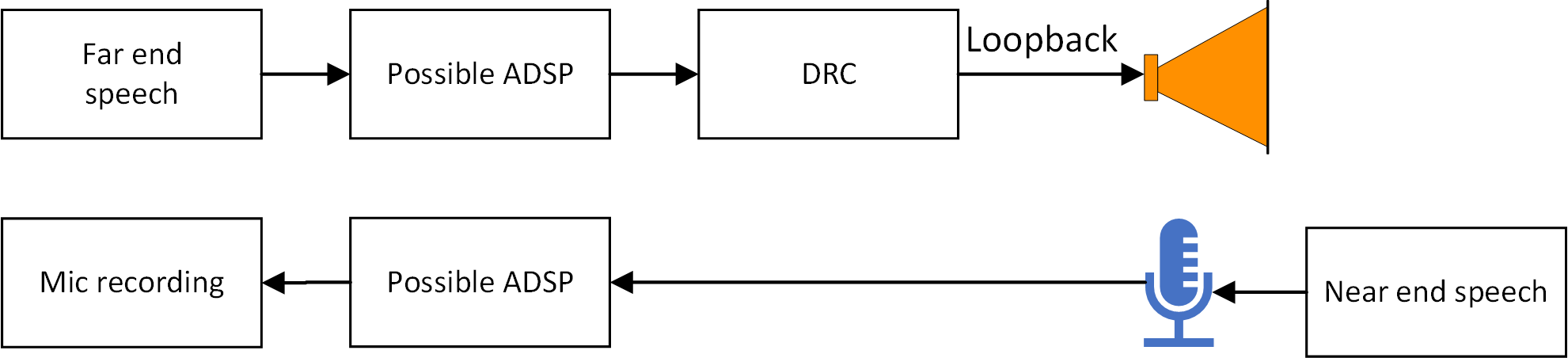}
    \caption{The custom recording application recorded the loopback and microphone signals.}
    \label{fig:recording}
\end{figure}

We use \emph{Amazon Mechanical Turk} as the crowdsourcing platform and wrote a custom HIT application that includes a custom tool that users download and execute to record the six scenarios described above. The dataset includes Microsoft Windows and Android devices. Each scenario includes the microphone and loopback signal (see Figure \ref{fig:recording}). Even though our application uses the WASAPI raw audio mode to bypass built-in audio effects, the PC can still include Audio DSP on the receive signal (e.g., equalization and Dynamic Range Compression); it can also include Audio DSP on the send signal, such as AEC and noise suppression.

For far end signals, we use both clean speech and real-world recordings. For clean speech far end signals, we use the speech segments from the Edinburgh dataset \cite{valentini-botinhao_speech_2016}. This corpus consists of short single speaker speech segments ($1$ to $3$ seconds). We used a long short term memory (LSTM) \cite{hochreiter_long_1997} based gender detector to select an equal number of male and female speaker segments. Further, we combined $3$ to $5$ of these short segments to create clips of length between $9$ and $15$ seconds in duration. Each clip consists of a single gender speaker. We create a gender-balanced far end signal source comprising of $500$ male and $500$ female clips. Recordings are saved at the maximum sampling rate supported by the device and in 32-bit floating point format; in the released dataset we down-sample to 48 kHz and 16-bit using automatic gain control to minimize clipping.

For noisy speech far end signals we use $2000$ clips from the near end single talk scenario. Clips are gender balanced to include an equal number of male and female voices.

For the far end single talk scenario, the clip is played back twice. This way, the echo canceller can be evaluated both on the first segment, when it has had minimal time to converge, and on the second segment, when the echo canceller has converged and the result is more indicative of a real call scenario.

For the double talk scenario, the far end signal is similarly played back twice, but with an additional silent segment in the middle, when only near end single talk occurs.

For near end speech, the users were prompted to read sentences from a TIMIT \cite{garofolo_darpa_1993} sentence list. Approximately 10 seconds of audio is recorded while the users are reading.

For track two (personalized AEC) we include 30 seconds of target speaker for each clip in the test set. In addition, the training and test set from the ICASSP 2022 Deep Noise Suppression Challenge track two \cite{dubey_icassp_2022} can be used.

\subsection{Synthetic dataset}
\label{sec:synth_data}
The second dataset provides 10,000 synthetic scenarios, each including single talk, double talk, near end noise, far end noise, and various nonlinear distortion scenarios. Each scenario includes a far end speech, echo signal, near end speech, and near end microphone signal clip. We use 12,000 cases (100 hours of audio) from both the clean and noisy speech datasets derived in \cite{reddy_interspeech_2020} from the LibriVox project\footnote{\url{https://librivox.org}} as source clips to sample far end and near end signals. The LibriVox project is a collection of public-domain audiobooks read by volunteers. \cite{reddy_interspeech_2020} used the online subjective test framework ITU-T P.808 to select audio recordings of good quality (4.3 $\leq$ MOS $\leq$ 5) from the LibriVox project. The noisy speech dataset was created by mixing clean speech with noise clips sampled from AudioSet \cite{gemmeke_audio_2017}, Freesound\footnote{\url{https://freesound.org}} and DEMAND \cite{thiemann_diverse_2013} databases at signal to noise ratios sampled uniformly from [0, 40] dB.

\begin{figure}
    \centering
    \includegraphics[width=0.5\textwidth]{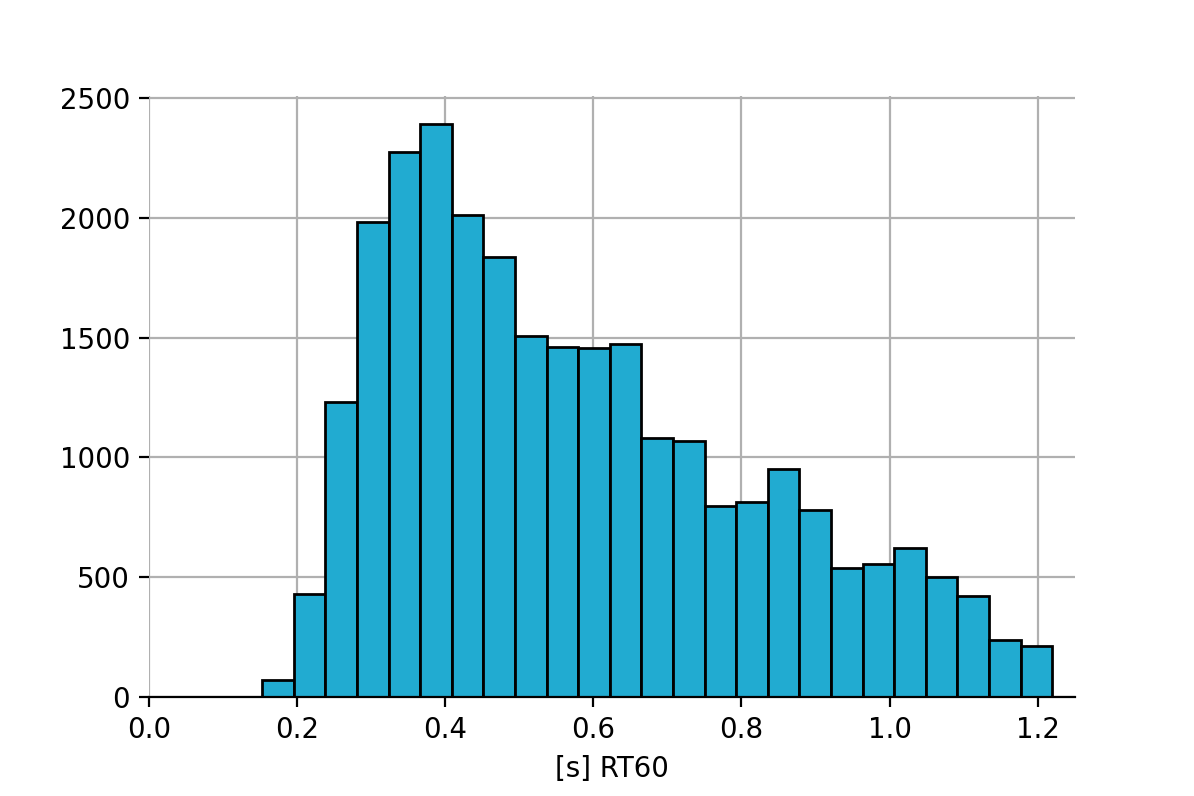} 
    \caption{Distribution of reverberation time (RT60) for the synthetic dataset.}
    \label{fig:rt60_synthetic}
\end{figure}

To simulate a far end signal, we pick a random speaker from a pool of 1,627 speakers, randomly choose one of the clips from the speaker, and sample 10 seconds of audio from the clip. For the near end signal, we randomly choose another speaker and take 3-7 seconds of audio which is then zero-padded to 10 seconds. The selected far end speakers were 71\% male, and 67\% of the near end speakers were male. To generate an echo, we convolve a randomly chosen room impulse response from a large Microsoft unreleased database with the far end signal. The room impulse responses are generated by using Project Acoustics technology\footnote{\url{https://www.aka.ms/acoustics}} and the RT60 ranges from 200 ms to 1200 ms. The distribution of RT60 is shown in Figure \ref{fig:rt60_synthetic}. In 80\% of the cases, the far end signal is processed by a nonlinear function to mimic loudspeaker distortion (the linear-to-nonlinear ratio is 0.25). For example, the transformation can be clipping the maximum amplitude, using a sigmoidal function as in \cite{lee_dnn-based_2015}, or applying learned distortion functions, the details of which we will describe in a future paper. This signal gets mixed with the near end signal at a signal-to-echo ratio uniformly sampled from -10 dB to 10 dB. The signal-to-echo ratio is calculated based on the clean speech signal (i.e., a signal without near end noise). The far end and near end signals are taken from the noisy dataset in 50\% of the cases. The first 500 clips can be used for validation as these have a separate list of speakers and room impulse responses. Detailed metadata information can be found in the repository.

\section{Test set}
\label{sec:data_test}

Two test sets are included, one at the beginning of the challenge and a blind test set near the end. Both consist of 800 real-world recordings, between ~30-45 seconds in duration. The datasets include the following scenarios that make echo cancellation more challenging:

\begin{itemize}
    \item Long- or varying delays, i.e., files where the delay between loopback and mic-in is atypically long or varies during the recording
    \item Strong speaker and/or mic distortions
    \item Stationary near end noise
    \item Non-stationary near end noise
    \item Recordings with audio DSP processing from the device, such as AEC or noise reduction
    \item Glitches, i.e., files with ``choppy" audio, for example, due to very high CPU usage
    \item Gain variations, i.e., recordings where far end level changes during the recording
  (\ref{ssec:real_data}), sampled randomly
\end{itemize}

\section{Baseline AEC Method}
\label{sec:model}
We adapt a noise suppression model developed in \cite{xia_weighted_2020} to the task of echo cancellation. 
Specifically, a recurrent neural network with gated recurrent units takes concatenated log power spectral features of the microphone signal and far end signal as input and outputs a spectral suppression mask. 
The short-time Fourier transform is computed based on 20ms frames with a hop size of 10 ms, and a 320-point discrete Fourier transform.
We use a stack of two gated recurrent unit layers, each of size 322 nodes, followed by a fully-connected layer with a sigmoid activation function. The model has 1.3 million parameters.
The estimated mask is point-wise multiplied by the magnitude spectrogram of the microphone signal to suppress the far end signal. 
Finally, to resynthesize the enhanced signal, an inverse short-time Fourier transform is used on the phase of the microphone signal and the estimated magnitude spectrogram.
We use a mean squared error loss between the clean and enhanced magnitude spectrograms. 
The Adam optimizer \cite{kingma_adam_2015} with a learning rate of 0.0003 is used to train the model. The model and the inference code are available in the challenge repository.\footnote{\url{https://github.com/microsoft/AEC-Challenge/tree/main/baseline/icassp2022}}

\section{Online subjective evaluation framework}
\label{sec:framework}
We have extended the open source P.808 Toolkit \cite{naderi_open_2020} with methods for evaluating echo impairments in subjective tests. We followed the \textit{Third-party Listening Test B} from ITU-T Rec. P.831 \cite{noauthor_itu-t_1998-1} and ITU-T Rec. P.832 \cite{noauthor_itu-t_2000} and adapted them to our use case as well as for the crowdsourcing approach based on the ITU-T Rec. P.808 \cite{noauthor_itu-t_2018} guidance. 

\begin{figure}[t]
    \centering
	\includegraphics[width=0.7\columnwidth]{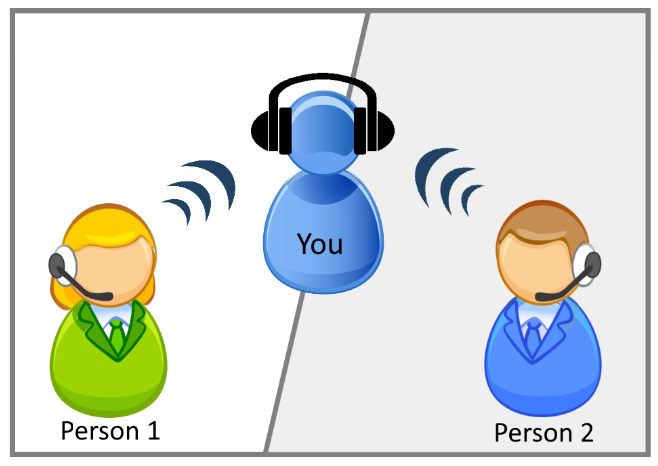}
	\caption{Double talk scenario in the Third Party Listening Test. The test participant (marked by "You") is positioned in the center of the communication.}
	\label{fig:double_talk}
\end{figure}

A third-party listening test differs from the typical listening-only tests (according to the ITU-T Rec. P.831) in the way that listeners hear the recordings from the \textit{center} of the connection rather in the former one in which the listener is positioned at one end of the connection \cite{noauthor_itu-t_1998-1} (see Figure \ref{fig:double_talk}). Thus, the speech material should be recorded by having this concept in mind.
During the test session, we use different combinations of single- and multi-scale Absolute Category Ratings depending on the speech sample under evaluation. We distinguish between single talk and double talk scenarios.
For the near end single talk, we ask for the overall quality. For the far end single talk and double talk scenario, we ask for an echo annoyance and for impairments of other degradations in two separate questions:

\begin{enumerate}
    \item How would you judge the degradation from the echo?
    \item How would you judge other degradations (noise, missing audio, distortions, cut-outs)
\end{enumerate}

Both impairments are rated on the degradation category scale (from 1: \textit{Very annoying}, to 5: \textit{Imperceptible}) to obtain degradation mean opinion scores (DMOS). Note that we do not use the Other degradation category for far end single talk for evaluating echo cancellation performance, since this metric mostly reflects the quality of the original far end signal. However, we have found that having this component in the questionnaire helps increase the accuracy of echo degradation ratings (when measured against expert raters). Without the Other category, raters can sometimes assign degradations due to noise to the Echo category  \cite{cutler_crowdsourcing_2021}.

\begin{figure}[t]
	\includegraphics[width=1\columnwidth]{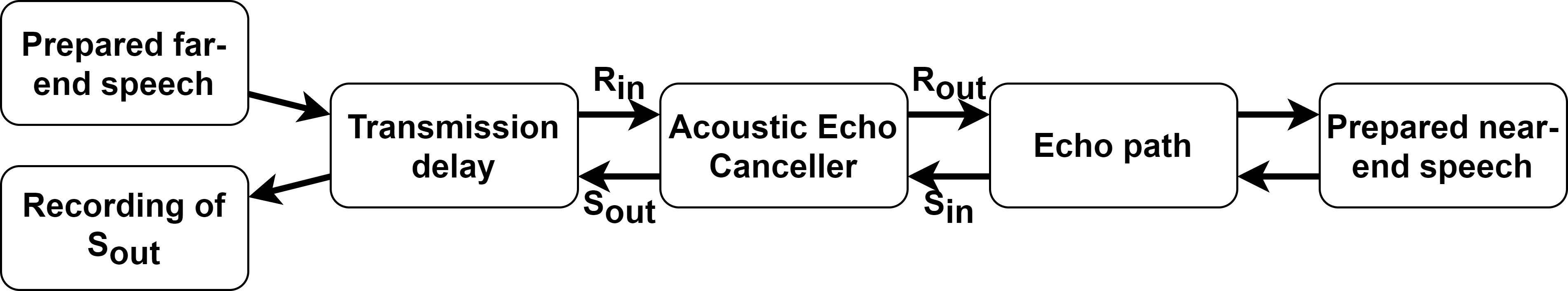}
	\caption{Echo canceller test set-up for Third Party Listening Test B according to the ITU-T Rec.P.831 (after \cite{noauthor_itu-t_1998-1}). $S$ is send and $R$ is receive.}
	\label{fig:P831}
\end{figure}

The setup illustrated in Figure \ref{fig:P831} is used to process all speech samples with all of the AECs under the study. To simplify the rating process for crowdworkers, we distinguished between near end and far end single talk as well as the double talk scenarios and tried to simulate them for the test participants. In the case of near end single talk we recorded the AEC output ($S_{out}$). For far end single talk, we added the output of the AEC ($S_{out}$) with a delay of 600ms to the loopback ($R_{in}$) signal, yielding $R_{in} + $ delayed $S_{out}$. For the listener, this simulates hearing the echo of their own speech (i.e., $R_{in}$ as an acoustic sidetone). For double talk the process is similar, but due to there being more speakers, simply adding the delayed AEC output ($S_{out}$) would cause confusion for the test participants. To mitigate this issue, the signals are played in stereo instead, with the loopback signal ($R_{in}$) played in one ear (i.e., acoustic sidetone) and the delayed output of the AEC ($S_{out}$) played in the other. Figure \ref{fig:double_talk} was used to illustrate the double talk scenario to crowdworkers.

For the far end single talk scenario, we evaluate the second half of each clip to avoid initial degradations from initialization, convergence periods, and initial delay estimation. For the double talk scenario, we evaluate the final third of the audio clip.

The subjective test framework is available at \url{https://github.com/microsoft/P.808}. A more detailed description of the test framework and its validation is given in \cite{cutler_crowdsourcing_2021}.

\section{Objective metric}
\label{sec:metric}
We have developed an objective perceptual speech quality metric called AECMOS. It can be used to stack rank different AEC methods based on MOS estimates with high accuracy. It is a neural network-based model that is trained using the ground truth human ratings obtained using our online subjective evaluation framework. The audio data used to train the AECMOS model is gathered from the numerous subjective tests that we conducted in the process of improving the quality of our AECs as well as the first two AEC challenge results. The performance of AECMOS on AEC models is given in Table~\ref{tab:AECMOS} compared with subjective human ratings on the 18 submitted models. A more detailed description of AECMOS is given in \cite{purin_aecmos_2022}. Sample code can be found on \url{https://aka.ms/aec-challenge}.

\begin{table}[tb]
\centering
\caption{AECMOS performance using Pearson's correlation coefficient (PCC), Spearman's rank correlation coefficient SRCC, and Kendall's Tau-b with a 95\% confidence interval \cite{kendall_new_1938}} 
\label{tab:AECMOS}
\begin{tabular}{l c c c c c} 
\toprule
Scenario & PCC  & SRCC & Tau-b 95\\
\midrule
Far end single talk echo DMOS & 0.99  & 0.92 & 0.83 \\ 
Near end single talk SIG MOS & 0.85 & 0.70 & 0.59\\
Near end single talk BAK MOS & 0.93 & 0.88 & 0.74\\
Double talk echo DMOS & 0.99 & 0.96 & 0.86\\ 
Double talk other DMOS & 0.96 & 0.91 & 0.85\\ 
\bottomrule
\end{tabular}
\end{table}

\section{Challenge metric}
\label{sec:challenge_metric}
The challenge performance is determined using the average of the five subjective scores described in Section \ref{sec:framework} and WAcc, all weighted equally; see Equation \eqref{eq:challenge_metric}, where $FE$ is far end single talk, $NE_{SIG}$ and $NE_{BAK}$ are P.835 SIG and BAK scores for near end single talk, $DT_{echo}$ is double talk echo, and $DT_{other}$ is double talk other.

\begin{figure*}
\begin{equation}
M = \frac{\frac{FE-1}{4} + \frac{NE_{SIG} - 1}{4} +\frac{NE_{BAK} - 1}{4} + \frac{DT_{echo}-1}{4}  + \frac{DT_{other}-1}{4} + WAcc}{6}
\label{eq:challenge_metric}
\end{equation}
\end{figure*}

\section{Results and analysis}
\label{sec:results}
The challenge had 20 entries, 17 for the non-personalized track and 3 for the personalized track. In addition, we included two internally developed models based on \cite{indenbom_deepvqe_2023}, labeled MS-1 and MS-2. We batched all submissions into three sets:

 \begin{itemize}
  \item Near end single talk files for a MOS test (NE ST MOS).
  \item Far end single talk files for an Echo and Other degradation DMOS test (FE ST Echo/Other DMOS).
  \item Double talk files for an Echo and Other degradation DMOS test (DT Echo/Other DMOS).
\end{itemize}

\begin{figure*}
    \centering
    \includegraphics[width=\textwidth]{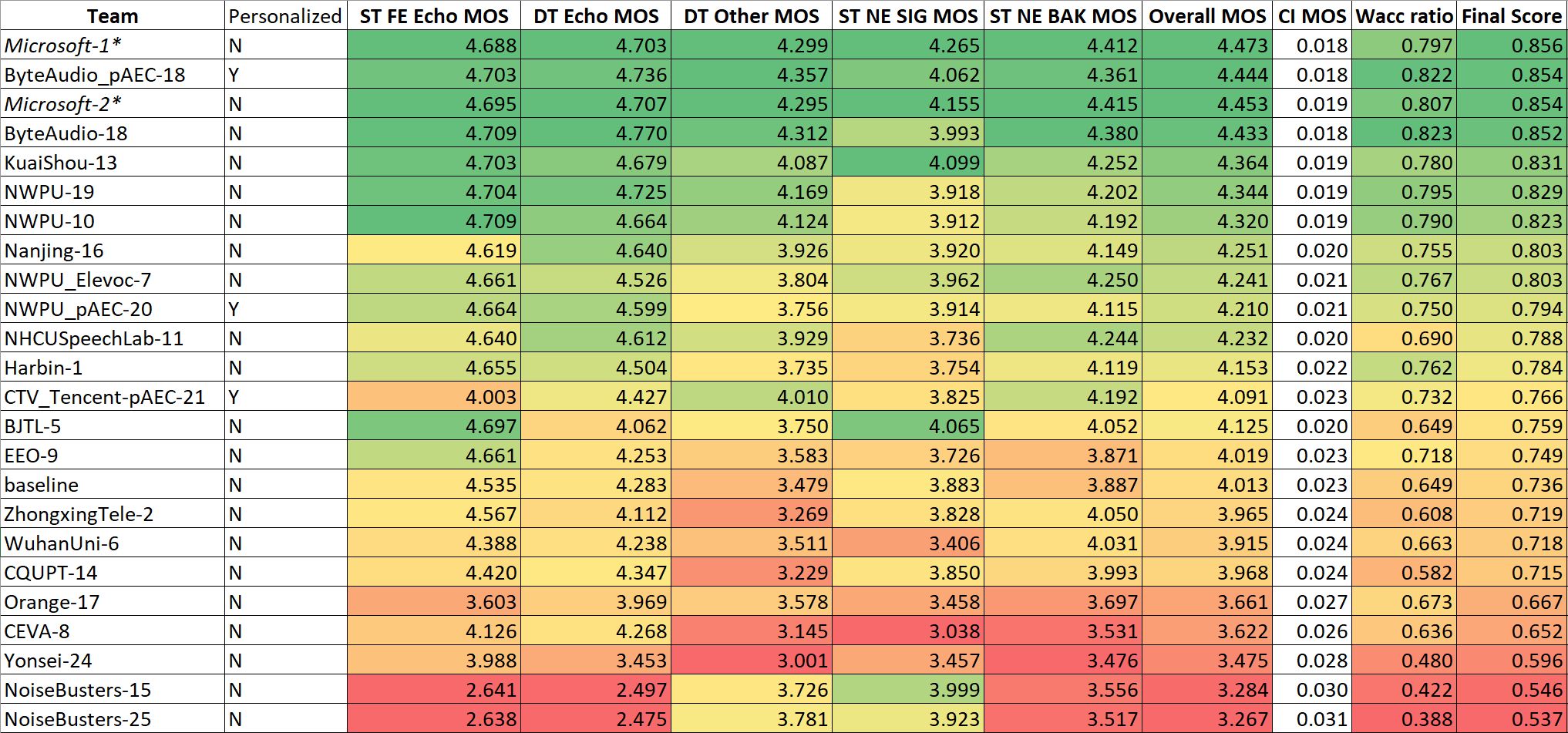}
    \caption{Challenge results. See Table \ref{tab:performance_metrics} for descriptions of the metrics. The Microsoft models were not officially entered in the challenge.}
    \label{fig:results}
\end{figure*}

The results are given in Figure \ref{fig:results}, and the analysis Of variance (ANOVA) for the top entries is given in Figure \ref{tab:anova}. The 2nd and 3rd places were tied. For the ties, the winners were selected using the lower complexity model.

A high-level comparison of the top-5 entries is given in Table \ref{tab:Top5}. Some observations are given below:

\begin{itemize}
    \item There is a PCC=-0.54 between the model size and the overall score. For this challenge, smaller models tend to outperform the larger models.
    \item There is a PCC=0.67 between the RTF and the overall score. More complex models tend to outperform the less complex models. 
    \item There is a PCC=0.10 between if the model was a hybrid and the overall score. Both hybrid and deep models perform well.
    \item There is a PCC=0.19 between the training dataset size and the overall score. Dataset size was not a significant factor in this challenge\footnote{For MS-1 and MS-2 we reported dataset size before augmentation.}.
    \item There is a PCC=0.49 between using additional datasets and the overall scale. Only one team added additional data (LibriSpeech \cite{panayotov_librispeech:_2015}), though they were the first-place team \cite{chen_progressive_2023}.
    \item The first-place entry showed that personalized AEC did increase performance, but only by a small amount (improving the final score by 0.002). 
\end{itemize}

\begin{table}[tb]
\centering
\caption{Performance metrics used in Figure \ref{fig:results}}
\label{tab:performance_metrics}
\begin{tabular}{l l} 
\toprule
Metric & Description \\
\midrule
ST FE Echo MOS &  Single talk far end echo MOS \\ 
DT Echo MOS & Double talk echo MOS \\
DT Other MOS & Other double talk distortion MOS \\
ST NE SIG MOS & Single talk near end speech signal MOS\\ 
ST NE BAK MOS & Single talk near end background MOS \\ 
WAcc & Word accuracy rate for speech recognition\\
\bottomrule
\end{tabular}
\end{table}


\begin{table*}
    \label{tab:anova}
    \centering
    \caption{ANOVA for the top challenge entries. The pair-wise p-values are shown for the lower triangular matrix only.}
    \begin{tabular}{lllllll}
    \toprule
                         & ByteAudio-18 & KuaiShou-13 & NWPU-19 & NWPU-10 & Nanjing-16 & NWPU\_Elevoc-7 \\
    \midrule
         KuaiShou-13     & 0.00 &  & & & & \\
         NWPU-19         & 0.00 & 0.34 & & & & \\
         NWPU-10         & 0.00 & 0.00 & 0.01 &  & & \\
         Nanjing-16      & 0.00 & 0.00 & 0.00 & 0.00 & & \\
         NWPU\_Elevoc-7  & 0.00 & 0.00 & 0.00 & 0.00 & 0.96 & \\
         NHCUSpeechLab-11 & 0.00 & 0.00 & 0.00 & 0.00 & 0.00 & 0.00 \\
    \bottomrule
    \end{tabular}
\end{table*}

\begin{table*}[tb]
\centering
\caption{Comparison of the top 5 teams} 
\label{tab:Top5}
\begin{tabular}{c c c c r c c c c} 
\toprule
Place & Track & Paper & Hybrid & Params & RTF & Dataset & Additional & Score \\
 &  &  &  &  &  & Hours & Datasets & \\
\hline
MS-1 & 1 & \cite{indenbom_deepvqe_2023} & N & 4.1 M & 0.42 & 950 &  & 0.856 \\ 
1 & 1 & \cite{chen_progressive_2023} & Y & 1.5 M & 0.28 & 2000 & LibriSpeech & 0.854 \\ 
MS-2 & 1 & \cite{indenbom_deepvqe_2023} & N & 7.5 M & 0.36 & 950 & & 0.854 \\ 
1 & 2 & \cite{chen_progressive_2023} & Y & 1.6 M & 0.31 & 2000 & LibriSpeech & 0.852 \\ 
2 & 1 & \cite{zhao_low-latency_2023} & N & 32.6 M & 0.19 & 605 & & 0.831 \\ 
3 & 1 & \cite{zhang_two-step_2023} & Y & 9.6 M & 0.35 & 3000 & & 0.829 \\ 
4 & 1 & \cite{sun_multi-task_2023} & Y & 3.8 M & 0.20 & 1400 & & 0.823 \\ 
5 & 1 & \cite{xu_tayloraecnet_2023} & N & 19.2 M & 0.22 & 600 & & 0.803 \\ 
\midrule
PCC to Score & & & 0.10 & -0.54 & 0.67 & 0.19 & 0.49 \\
\bottomrule
\end{tabular}
\end{table*}





\subsection{Performance comparison of the ICASSP 2022 and 2023 AEC Challenge}

To compare the winning model performance from the ICASSP 2022 AEC Challenge to models from this year's challenge,  we apply the top-scoring model MS-1 on 2022 AEC Challenge data and use the online subjective evaluation framework to compare the results. Table \ref{tab:compare2022} shows that MS-1 is statistically the same as the 2022 AEC Challenge winner \cite{zhang_multi-scale_2022}, even though the algorithmic latency + buffering latency for MS-1 is 20ms and for \cite{zhang_multi-scale_2022} 40ms. In studies with the CRUSE \cite{braun_towards_2021} noise suppression model which MS-1 is based on, changing the frame size from 20ms to 40ms increased DNSMOS OVRL by 0.1. In addition, changing the frame size of MS-1 from 20ms to 10ms decreased DNSMOS OVRL by 0.07. Therefore, we conclude that MS-1 should be significantly better than \cite{zhang_multi-scale_2022} if that model also had an algorithmic + buffering latency of 20ms.

\begin{table}[tb]
    \centering
    \caption{Comparison of top performing models from 2022 and 2023 challenge.}
    \begin{tabular}{lcc}
    \toprule
        & MS-1 & 2022  Winner \cite{zhang_multi-scale_2022} \\
        \hline 
        Single talk far end echo MOS & 4.65 & 4.66\\ 
        Double talk echo MOS & 4.68 & 4.68\\ 
        Double talk other MOS & 4.26 & 4.28\\ 
        Algorithmic + buffering latency & 20ms & 40ms\\
    \bottomrule
    \end{tabular}
    \label{tab:compare2022}
\end{table}

\section{Conclusions}
\label{sec:conclusions}
This latest AEC challenge induced lower algorithmic latency + buffering latency requirements and added a personalized track. The performance of the top models is exceptional, though it shows there is still a lot of headroom to improve, especially in the double talk other, single  near end, and WAcc metrics (see Table \ref{tab:improvement2}). We are optimistic that the personalized enrollment data can improve these areas much more than was shown in this challenge, which is a good area for future research. In addition, even lower latency requirements are needed for a telecommunication system to achieve end-to-end latencies of less than 50ms, which is the just-noticeable difference when latency impacts conversations \cite{kitawaki_pure_1991}. End-to-end latencies significantly above 50ms have been shown to be correlated to lower participation in group meetings \cite{bothra_dont_2023}. To achieve this goal the algorithmic latency + buffering latency should be less than 5ms, which is another good area for future work.

\begin{table}
    \centering
    \caption{Amount of improvement remaining to get excellent quality rated speech or perfect WAcc on this challenge.}
    \begin{tabular}{lc}
    \toprule
        Area & Headroom \\
        \hline 
        Single talk far end echo & 0.30 \\
        Double talk echo & 0.26 \\
        Double talk other & 0.64 \\
        Single talk near end SIG & 0.74 \\
        Single talk near end BAK & 0.64  \\
        Overall & 0.56 \\
        \hline
        WAcc & 0.18 \\
    \bottomrule
    \end{tabular}
    \label{tab:improvement2}
\end{table}

\FloatBarrier 
\bibliographystyle{IEEEbib}
\bibliography{IC3-AI}
\end{document}